\documentclass[reqno,10pt,a4paper]{amsart}

\usepackage{amssymb,mathptmx,cite,array,setspace,geometry,enumitem}
\usepackage{graphicx}
\usepackage{tikz}
\usetikzlibrary{shapes.geometric} 

\numberwithin{equation}{section}

\newcolumntype{C}{>{$}c<{$}} 

\allowdisplaybreaks

\newcommand{\alg}[1]{\mathfrak{#1}}
\newcommand{\vir}{\alg{Vir}}

\newcommand{\func}[2]{#1 \left( #2 \right)}

\newcommand{\brac}[1]{\left( #1 \right)}
\newcommand{\sqbrac}[1]{\left[ #1 \right]}
\newcommand{\set}[1]{\left\{ #1 \right\}}

\newcommand{\ZZ}{\mathbb{Z}}

\newcommand{\CC}{\mathbb{C}}

\newcommand{\bra}[1]{\bigl\langle #1 \bigr\rvert}
\newcommand{\ket}[1]{\bigl\lvert #1 \bigr\rangle}
\newcommand{\braket}[2]{\bigl\langle #1 \bigr\rvert \bigl. #2 \bigr\rangle}
\newcommand{\bracket}[3]{\bigl\langle #1 \bigr\rvert #2 \bigl\lvert #3 \bigr\rangle} 

\newcommand{\IrrMod}[1]{\mathcal{L}_{#1}}
\newcommand{\VerMod}[1]{\mathcal{V}_{#1}}
\newcommand{\QuotMod}[1]{\mathcal{Q}_{#1}}
\newcommand{\StagMod}[1]{\mathcal{S}_{#1}}

\newcommand{\eqnref}[1]{Equation~\eqref{#1}}
\newcommand{\eqnDref}[2]{Equations~\eqref{#1} and \eqref{#2}}
\newcommand{\secref}[1]{Section~\ref{#1}}

\newcommand{\cft}{conformal field theory}

\newcommand{\lcft}{logarithmic conformal field theory}
\newcommand{\lcfts}{logarithmic conformal field theories}
\newcommand{\ope}{operator product expansion}
\newcommand{\opes}{operator product expansions}
\newcommand{\hws}{highest weight state}

\newcommand{\hwm}{highest weight module}

\newcommand{\fuse}{\mathbin{\times}}

\DeclareMathOperator{\id}{id}

\newcommand{\ahol}[1]{\overline{#1}}

\newcommand{\bvac}{\mathbf{0}}
\newcommand{\bT}{\mathbf{T}}
\newcommand{\bt}{\mathbf{t}}
\newcommand{\bbT}{\ahol{\bT}}
\newcommand{\bbt}{\ahol{\bt}}

\newcommand{\bL}{\mathbf{L}}
\newcommand{\bbL}{\ahol{\bL}}

\newcommand{\bcS}{\pmb{\mathcal{S}}}
\newcommand{\bcT}{\pmb{\mathcal{T}}}

\newcommand{\mbcS}{\bcS_{\mspace{-12mu} \raisebox{-2pt}{$\circ$}}}

\newcommand{\bOmega}{\pmb{\Omega}}
\newcommand{\bomega}{\pmb{\omega}}
\newcommand{\bbomega}{\ahol{\bomega}}
\newcommand{\bchi}{\ahol{\chi}}

\newcommand{\bdb}{b}
\newcommand{\bbb}{\ahol{\bdb}}
\newcommand{\bbeta}{\ahol{\beta}}
\newcommand{\bh}{\ahol{h}}
\newcommand{\bn}{\ahol{n}}
\newcommand{\bp}{\ahol{p}}
\newcommand{\bz}{\ahol{z}}

\newcommand{\bx}{\mathbf{x}}
\newcommand{\by}{\mathbf{y}}
\newcommand{\bU}{\mathbf{U}}
\newcommand{\bV}{\mathbf{V}}
\newcommand{\bby}{\ahol{\by}}
\newcommand{\bbU}{\ahol{\bU}}
\newcommand{\bbV}{\ahol{\bV}}

\newcommand{\pa}{\partial}
\newcommand{\bpa}{\ahol{\pa}}

\begin{document}

\title{Non-Chiral Logarithmic Couplings for the Virasoro Algebra}

\author[D Ridout]{David Ridout}

\address[David Ridout]{
Department of Theoretical Physics \\
Research School of Physics and Engineering;
and
Mathematical Sciences Institute;
Australian National University \\
Canberra, ACT 0200 \\
Australia
}

\email{david.ridout@anu.edu.au}

\date{\today}

\begin{abstract}
This note initiates the study of what we call \emph{non-chiral staggered Virasoro modules}, indecomposable modules on which two copies of the Virasoro algebra act with the zero-modes $\bL_0$ and $\bbL_0$ acting non-semisimply.  This is motivated by the ``puzzle'' recently reported in \cite{VasPuz11} involving a non-standard measured value, meaning that the value is not familiar from chiral studies, for the ``$b$-parameter'' (logarithmic coupling) of a $c=0$ bulk \cft{}.  Here, an explanation is proposed by introducing a natural family of bulk modules and showing that the only consistent, non-standard logarithmic coupling that is distinguished through structure is that which was measured.  This observation is shown to persist for general central charges and a conjecture is made for the values of certain non-chiral logarithmic couplings.
\end{abstract}

\maketitle

\onehalfspacing

\section{Introduction and Background} \label{secIntro}

A recent advance in our understanding of critical phenomena in two dimensions is that the non-local observables, such as crossing probabilities and fractal dimensions, of statistical lattice models are not described in the scaling limit by rational \cft{}, but rather by \emph{logarithmic} \cft{}.  The difference amounts to the presence of correlation functions with logarithmic singularities instead of just poles and root singularities \cite{RozQua92}.  At the level of states, this turns out to be essentially equivalent to having a zero-mode, usually the Virasoro mode $L_0 \in \vir$, acting non-diagonalisably \cite{GurLog93}.

The case of models with scaling limit of central charge $c=0$ is particularly interesting because it includes such paradigms as percolation and the self-avoiding walk (dilute polymers), even though standard theories at this central charge are trivial.  Indeed, the literature devoted to making sense of $c=0$ \lcfts{} has grown rather dramatically in recent years.  Our interest here relates to a recent observation of \cite{VasPuz11} concerning the value of a certain parameter that was measured (approximately) for both percolation and dilute polymers when periodic boundary conditions are imposed.  The result, which is characteristic of the bulk $c=\ahol{c}=0$ theories, was unexpected as it differed from that predicted through chiral considerations (pertaining to the boundary theory).  Our purpose here is to explain that the measured bulk value has a perfectly good structural explanation which mirrors that of the boundary values.  More precisely, this value follows directly from a very natural Ansatz for the bulk module structure.  We remark that the structure of bulk $c=0$ theories has already been considered in the literature \cite{SalGL106,SalSU207,GabMod11,RunLog12}, but only from the perspective of extended algebras.  It would be very interesting to compare the structures obtained there with the Virasoro structures proposed below.

To be more specific, the ``certain parameter'' alluded to above is a complex number which parametrises the isomorphism classes of certain types of indecomposable $\vir$-modules called staggered modules \cite{RohRed96,RidSta09}.  Initially labelled by $\beta$, such parameters were first introduced in \cite{GabInd96} for a small set of modules with $c=-2$ and \mbox{$c=-7$}.  Shortly thereafter, similar parameters $b$ were identified \cite{GurCTh99} in \opes{} at $c=0$ and referred to as \emph{anomaly numbers}.  Many further examples of these numbers were computed in \cite{EbeVir06}, though the definition used there failed to give reproducible results in all but the simplest cases.  This was corrected in \cite{RidPer07}, where an invariant definition for $\beta$ was given for certain physically relevant classes of staggered $\vir$-modules (see \cite{RidSta09} for general staggered modules).  The invariant $\beta$ were christened \emph{logarithmic couplings} in \cite{RidPer07}, although the terms \emph{beta-invariants} \cite{RidSta09} and \emph{indecomposability parameters} \cite{DubCon10,VasInd11} have also been used since.

A staggered $\vir$-module is defined in \cite{RidSta09} to be an extension of a \hwm{} by another \hwm{} upon which the Virasoro mode $L_0$ acts non-diagonalisably.  This zero-mode then possesses a rank $2$ Jordan cell of minimal conformal dimension $h$ spanned by two states $\ket{\psi}$ and $\ket{\phi} = \brac{L_0 - h} \ket{\psi}$.  It is easy to check that $\ket{\phi}$ is a singular vector and that the logarithmic coupling
\begin{equation} \label{eqn:DefBeta}
\beta = \braket{\phi}{\psi}
\end{equation}
is independent of the choice of Jordan partner $\ket{\psi}$ \cite{RidPer07}.  As one typically has an infinite number of staggered $\vir$-modules in a \lcft{}, there are usually infinitely many logarithmic couplings to compute.  We mention that for any given staggered module, the actual value of $\beta$ depends upon the chosen normalisation for the singular vector $\ket{\phi}$ relative to the \hws{} $\ket{\xi}$ from which it is descended.\footnote{We assume here that $\ket{\phi}$ is not proportional to $\ket{\xi}$ and that $\braket{\xi}{\xi} = 1$.  When $\ket{\phi} = \ket{\xi}$, staggered modules are unique (when they exist) \cite{RidSta09}, so one does not define $\beta$ in this situation.}  Unfortunately, there are two competing conventions that are used for this normalisation:  If $\ket{\phi}$ is a grade $n>0$ descendant of $\ket{\xi}$, then one writes either
\begin{equation} \label{eqn:NormSV}
\ket{\phi} = \brac{L_{-1}^n + \cdots} \ket{\xi} \qquad \text{or} \qquad \ket{\phi} = \brac{L_{-n} + \cdots} \ket{\xi}.
\end{equation}
The first has some theoretical advantages \cite{RidLog07,RidSta09}.  In particular, the coefficient of $L_{-1}^n$ (but not $L_{-n}$) is known to be non-zero \cite{AstStr97} (when $\ket{\phi}$ belongs to a Verma module).  Moreover, we must assume in the second (but need not in the first) that the omitted monomials in the $L_{-m}$ are ordered in some standard manner, for example in Poincar\'{e}-Birkhoff-Witt order.  Nevertheless, it is convenient (and customary) to use the second convention when studying the particular $c=0$ examples which will mostly occupy us here.

In the chiral setting, the $c=0$ examples that have received the most attention concern those staggered modules which contain the vacuum state $\ket{0}$.  For this central charge, the state $\ket{T} = L_{-2} \ket{0}$ that corresponds to the energy-momentum tensor is singular and so one introduces a logarithmic partner $\ket{t}$ such that $\brac{L_0 - 2} \ket{t} = \ket{T}$.  The logarithmic coupling $\beta = \braket{T}{t}$ is then precisely the anomaly number $b$ of \cite{GurCTh99}.  In \cite{GurCon02,GurCon04}, \ope{} computations resulted in a heuristic derivation of the values of $b$ relevant for a $c=0$ \emph{chiral} \lcft{}:  $b = -\tfrac{5}{8}$ and $b = \tfrac{5}{6}$ (assuming the second normalisation of \eqref{eqn:NormSV}).  These values were recovered in \cite{EbeVir06} at the level of $\vir$-modules by explicitly constructing fusion products of $c=0$ irreducibles and \cite{RidPer07} then used this to identify $b = -\tfrac{5}{8}$ as characterising percolation and $b = \tfrac{5}{6}$ as characterising dilute polymers.  Moreover, it was explained in \cite{RidLog07}, and then proven in \cite{RidSta09}, that structural reasons forced these to be the only possible values for (self-contragredient) staggered modules containing the vacuum.  These values have even been confirmed directly at the level of the lattice for both percolation and dilute polymers \cite{DubCon10}.

The above results concerning $b$ pertain either to boundary or chiral \cft{} at $c=0$.  Nevertheless, recent measurements \cite{VasPuz11} of the lattice approximations to $b$ for the \emph{non-chiral}, meaning bulk, $c=\ahol{c}=0$ theories describing percolation and dilute polymers, came as a surprise because they appear to converge to $-5$.  This limit was confirmed using two different theoretical arguments:  The first is a remarkable heuristic \cite{VasInd11} by which one recovers logarithmic operator product expansions by taking carefully chosen limits of ``generic'' non-logarithmic expansions as the central charge and conformal dimensions tend to their required values.  The second is a Coulomb gas approach which again relies essentially on a similar limiting process.  In what follows, we generalise the structural arguments of \cite{RidLog07,RidSta09} from staggered $\vir$-modules to their natural $\vir \oplus \vir$-module analogues, that is, from the chiral to the non-chiral setting, to explain the measurement $b = -5$ from a representation-theoretic perspective.  This allows us to isolate the structural feature that seems to be responsible for this measurement and also to propose potential generalisations to other central charges.

We conclude this historical survey with a brief comment on the physical relevance of the logarithmic couplings $\beta$ that are being studied.  As is well known, they appear in \opes{} and, in particular, correlation functions where they play the role of two-point constants for logarithmic partner fields.  As such, they are fundamental quantities that need computing in any \lcft{}.  What seems to be not so well known, however, is that the $\beta$ one computes at the level of modules is \emph{not} precisely the same as that which appears in two-point functions.  The difference is a combinatorial proportionality factor which derives from relating $\bra{\phi}$ in \eqref{eqn:DefBeta} to the singular field $\func{\phi}{z}$.  We refer to \cite[App.\ A]{RidLog07} for a simple example illustrating this subtlety (see also \cite{GaiLat12}) and remark that the seeming obscurity of the difference between the representation-theoretic $\beta$ and the correlator $\beta$ probably stems from the fact that for the $c=0$ vacuum modules (where $\beta$ is customarily denoted by $b$), this proportionality factor is $1$.

\section{A Non-Chiral Staggered Module}

The indecomposable $\vir \oplus \vir$-module studied in \cite{VasPuz11} has the (non-chiral) vacuum module $\QuotMod{1,1} \otimes \QuotMod{1,1}$ as a submodule.  Here, $\QuotMod{1,1}$ is the $c=0$ $\vir$-module\footnote{Until further notice, all $\vir$-modules will be assumed to be $c=0$.} obtained from the vacuum Verma module $\VerMod{0}$ by setting the Verma submodule $\VerMod{1}$, generated by the singular vector $L_{-1} \ket{0}$, to zero.  That is, $\QuotMod{1,1} = \VerMod{0} / \VerMod{1}$.  In general, we will denote by $\VerMod{h}$ the Verma $\vir$-module generated by a \hws{} of conformal dimension $h$ and by $\QuotMod{r,s}$ the quotient $\VerMod{h_{r,s}} / \VerMod{h_{r,s} + rs}$, where $h_{r,s}$ denotes the conformal dimension of the $\brac{r,s}$-entry in the Kac table.  The chiral vacuum module $\QuotMod{1,1}$ is itself reducible, with submodule $\IrrMod{2}$ and quotient $\QuotMod{1,1} / \IrrMod{2} = \IrrMod{0}$, where $\IrrMod{h}$ denotes the irreducible $\vir$-module generated by a \hws{} of conformal dimension $h$.  We illustrate this schematically with the aid of a structure diagram:
\begin{equation}
\parbox[c]{0.12\textwidth}{
\begin{tikzpicture}[auto,thick,
	nom/.style={circle,draw=black!20,fill=black!20,inner sep=2pt}
	]
\node (q1) at (0,1) {$\IrrMod{0}$};
\node (s1) at (0,-1) {$\IrrMod{2}$};
\node at (-1,0) [nom] {$\QuotMod{1,1}$};
\draw [->] (q1) to (s1);
\end{tikzpicture}
}.
\end{equation}
This indicates the decomposition of a module into its irreducible composition factors, with arrows indicating (schematically) the action of the algebra.  For example, we see that $\IrrMod{2}$ above is a submodule, because the arrow points towards it, whereas $\IrrMod{0}$ is only a (sub)quotient, because the arrow points away from it.  It follows that the indecomposable structure of the non-chiral vacuum module $\QuotMod{1,1} \otimes \QuotMod{1,1}$ is given by the following structure diagram:
\begin{equation}
\parbox[c]{0.38\textwidth}{
\begin{center}
\begin{tikzpicture}[auto,thick,
	nom/.style={ellipse,minimum height=8mm,draw=black!20,fill=black!20,inner sep=2pt}
	]
\node (top) at (0,1.5) [] {$\IrrMod{0} \otimes \IrrMod{0}$};
\node (left) at (-2,0) [] {$\IrrMod{2} \otimes \IrrMod{0}$};
\node (right) at (2,0) [] {$\IrrMod{0} \otimes \IrrMod{2}$};
\node (bot) at (0,-1.5) [] {$\IrrMod{2} \otimes \IrrMod{2}$};
\node at (0,0) [nom] {$\QuotMod{1,1} \otimes \QuotMod{1,1}$};
\draw [->] (top) to (left);
\draw [->] (top) to (right);
\draw [->] (left) to (bot);
\draw [->] (right) to (bot);
\end{tikzpicture}
\end{center}
}
.
\end{equation}
The space of singular vectors in $\QuotMod{1,1} \otimes \QuotMod{1,1}$ is therefore four-dimensional, and we shall choose basis vectors for this space in the obvious manner:
\begin{equation} \label{eqn:SVs}
\begin{split}
\ket{\bvac} &\equiv \ket{0} \otimes \ket{0}, \\
\ket{\bT} &\equiv \ket{T} \otimes \ket{0} = L_{-2} \ket{0} \otimes \ket{0} = \bL_{-2} \ket{\bvac}, \\
\ket{\bbT} &\equiv \ket{0} \otimes \ket{T} = \ket{0} \otimes L_{-2} \ket{0} = \bbL_{-2} \ket{\bvac}, \\
\ket{\bT \bbT} &\equiv \ket{T} \otimes \ket{T} = L_{-2} \ket{0} \otimes L_{-2} \ket{0} = \bL_{-2} \bbL_{-2} \ket{\bvac}.
\end{split}
\end{equation}
Here, $\bL_n$ and $\bbL_n$ denote $L_n \otimes \id$ and $\id \otimes L_n$, respectively.\footnote{In general, we shall use bold type to indicate non-chiral states and operators.}

To obtain what we shall refer to as a non-chiral staggered module, we introduce Jordan partner states $\ket{\bt}$ and $\ket{\bbt}$ to $\ket{\bT}$ and $\ket{\bbT}$ with (generalised) conformal dimensions $\brac{2,0}$ and $\brac{0,2}$, respectively:
\begin{subequations}
\begin{align}
\brac{\bL_0 - 2} \ket{\bt} &= \ket{\bT}, & \brac{\bbL_0 - 2} \ket{\bbt} &= \ket{\bbT}.
\intertext{This is physically reasonable as neither $\func{\bT}{z}$ nor $\func{\bbT}{\bz}$ have a conjugate field among those corresponding to the states of $\QuotMod{1,1} \otimes \QuotMod{1,1}$.  A necessary condition for the locality of the two-point functions in any non-chiral \cft{} is that $\bL_0 - \bbL_0$ be diagonalisable \cite{GabLoc99}, hence we may assume that}
\bbL_0 \ket{\bt} &= \ket{\bT}, & \bL_0 \ket{\bbt} &= \ket{\bbT}.
\end{align}
\end{subequations}
We will declare, for now, that there are no linear dependencies among the descendants of $\ket{\bt}$ and $\ket{\bbt}$.  Mathematically, this means that we are considering an indecomposable $\vir \oplus \vir$-module $\bcS'$ with submodule $\QuotMod{1,1} \otimes \QuotMod{1,1}$ and quotient
\begin{equation}
\frac{\bcS'}{\QuotMod{1,1} \otimes \QuotMod{1,1}} \cong \brac{\VerMod{2} \otimes \VerMod{0}} \oplus \brac{\VerMod{0} \otimes \VerMod{2}}.
\end{equation}
In other words, the descendants of $\ket{\bt}$ and $\ket{\bbt}$ form independent, non-chiral Verma modules, modulo the states of $\QuotMod{1,1} \otimes \QuotMod{1,1}$.

We should emphasise that the existence of this \emph{non-chiral staggered module} $\bcS'$ has not been proven.  While this can be seen from straight-forward generalisations of the arguments in \cite[Sec.~4--6]{RidSta09}, it is far beyond the scope of the note to address this question in the generality it deserves; we hope to return to it in a future publication.  In fact, these arguments allow us to assert something stronger:  That there is a two-parameter family of (isomorphism classes of) such $\bcS'$ which may be distinguished by two logarithmic couplings $\bdb$ and $\bbb$, defined by
\begin{equation} \label{eqn:NCBetas}
\bL_2 \ket{\bt} = \bdb \ket{\bvac}, \qquad \bbL_2 \ket{\bbt} = \bbb \ket{\bvac}.
\end{equation}
These couplings are clearly invariants of $\bcS'$ in that they do not depend upon the choice of $\ket{\bt}$ and $\ket{\bbt}$.  The quantity $\bdb$ appears to be precisely what was measured in \cite{VasPuz11} with the result $-5$.

The strategy now \cite{RidLog07} is to analyse whether one can introduce a linear dependence in the descendants of $\ket{\bt}$ and $\ket{\bbt}$, without affecting any of the states of the submodule $\QuotMod{1,1} \otimes \QuotMod{1,1}$, and check if imposing this linear dependence leads to any constraints on $\bdb$ and $\bbb$.  Mathematically, we are asking for the existence of submodules $\bcT' \subset \bcS'$ whose intersection with $\QuotMod{1,1} \otimes \QuotMod{1,1}$ is trivial and questioning whether there are submodules of this type that only exist for certain $\bdb$ and $\bbb$.  If we find such a submodule that only exists when $\bdb = -5$, then we have a seemingly strong candidate, namely $\bcS = \bcS' / \bcT'$, for explaining the measurement of \cite{VasPuz11}.

Before embarking on this quest, we find it useful to draw (a part of) the structure diagram for $\bcS'$.  This requires identifying the irreducible composition factors that correspond to the singular vectors of the non-chiral modules $\QuotMod{1,1} \otimes \QuotMod{1,1}$, $\VerMod{2} \otimes \VerMod{0}$ and $\VerMod{0} \otimes \VerMod{2}$.  The former has four singular vectors as we have seen, but the latter two have infinitely many.  For example, $\VerMod{2} \otimes \VerMod{0}$ has singular vectors of conformal dimension $\brac{h,\bh}$, where $h \in \set{2,5,7,12,15,\ldots}$ and $\bh \in \set{0,1,2,5,7,12,15,\ldots}$.  We will therefore only indicate a few of these, denoting (for clarity) the non-chiral composition factors by their conformal dimensions $\brac{h,\bh}$:
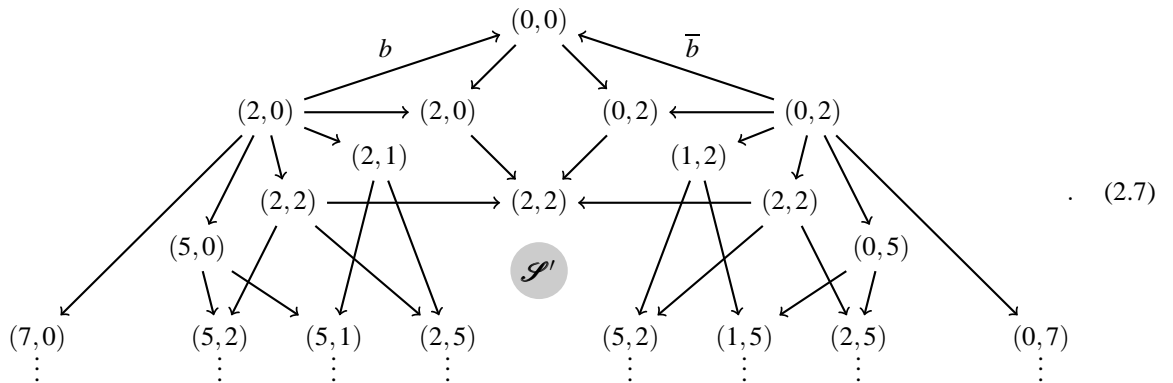
\begin{equation} \label{pic:S'}
\parbox[c]{0.9\textwidth}{
\begin{center}
\begin{tikzpicture}[auto,thick,scale=0.6,
	nom/.style={circle,draw=black!20,fill=black!20,inner sep=2pt}
	]
\node (top) at (0,0) [] {$\brac{0,0}$};
\node (left) at (-2,-2) [] {$\brac{2,0}$};
\node (right) at (2,-2) [] {$\brac{0,2}$};
\node (bot) at (0,-4) [] {$\brac{2,2}$};
\node (l20) at (-6,-2) [] {$\brac{2,0}$};
\node (l21) at (-3.5,-3) [] {$\brac{2,1}$};
\node (l22) at (-5.5,-4) [] {$\brac{2,2}$};
\node (l25) at (-2,-7) [] {$\brac{2,5}$};
\node (l50) at (-7.5,-5) [] {$\brac{5,0}$};
\node (l51) at (-4.5,-7) [] {$\brac{5,1}$};
\node (l52) at (-7,-7) [] {$\brac{5,2}$};
\node (l70) at (-11,-7) [] {$\brac{7,0}$};
\node (r02) at (6,-2) [] {$\brac{0,2}$};
\node (r12) at (3.5,-3) [] {$\brac{1,2}$};
\node (r22) at (5.5,-4) [] {$\brac{2,2}$};
\node (r52) at (2,-7) [] {$\brac{5,2}$};
\node (r05) at (7.5,-5) [] {$\brac{0,5}$};
\node (r15) at (4.5,-7) [] {$\brac{1,5}$};
\node (r25) at (7,-7) [] {$\brac{2,5}$};
\node (r07) at (11,-7) [] {$\brac{0,7}$};
\node at (0,-5.5) [nom] {$\bcS'$};
\node at (-11,-7.5) [] {$\vdots$};
\node at (-7,-7.5) [] {$\vdots$};
\node at (-4.5,-7.5) [] {$\vdots$};
\node at (-2,-7.5) [] {$\vdots$};
\node at (11,-7.5) [] {$\vdots$};
\node at (4.5,-7.5) [] {$\vdots$};
\node at (7,-7.5) [] {$\vdots$};
\node at (2,-7.5) [] {$\vdots$};
\draw [->] (top) to (left);
\draw [->] (top) to (right);
\draw [->] (left) to (bot);
\draw [->] (right) to (bot);
\draw [->] (l20) to (left);
\draw [->] (l22) to (bot);
\draw [->] (l20) to (l21);
\draw [->] (l20) to (l22);
\draw [->] (l21) to (l25);
\draw [->] (l22) to (l25);
\draw [->] (l20) to (l50);
\draw [->] (l20) to (l70);
\draw [->] (l22) to (l52);
\draw [->] (l50) to (l51);
\draw [->] (l50) to (l52);
\draw [->] (l21) to (l51);
\draw [->] (r02) to (right);
\draw [->] (r22) to (bot);
\draw [->] (r02) to (r12);
\draw [->] (r02) to (r22);
\draw [->] (r12) to (r52);
\draw [->] (r22) to (r52);
\draw [->] (r02) to (r05);
\draw [->] (r02) to (r07);
\draw [->] (r22) to (r25);
\draw [->] (r05) to (r15);
\draw [->] (r05) to (r25);
\draw [->] (r12) to (r15);
\draw [->] (l20) to node {$\bdb$} (top);
\draw [->] (r02) to node [swap] {$\bbb$} (top);
\end{tikzpicture}
\end{center}
}
.
\end{equation}
Despite appearances, this diagram is in fact a simplified version --- we have omitted all arrows corresponding to the action of the positive-mode subalgebra, except for those labelled by $\bdb$ and $\bbb$ which account for the actions described in \eqref{eqn:NCBetas}.  Indeed, our quest essentially involves determining whether certain of these omitted arrows are present or not (for certain values of $\bdb$ and $\bbb$).

We begin by noting that the maximal proper submodule of $\VerMod{2} \otimes \VerMod{0}$ is generated by four singular vectors with conformal dimensions $\brac{h,\bh} = \brac{5,0}$, $\brac{7,0}$, $\brac{2,1}$ and $\brac{2,2}$.  We will analyse the submodules generated by their lifts $\ket{\chi_{h,\bh}}$ in $\bcS'$.\footnote{We expect, again based on \cite{RidSta09}, that the non-generating singular vectors will not lead to constraints on $\bdb$ and $\bbb$.}  These are defined to be states of $\bcS'$ which are mapped to the appropriate singular vectors upon quotienting by $\QuotMod{1,1} \otimes \QuotMod{1,1}$.  It follows that the action of $\bL_n$ and $\bbL_n$ for $n>0$, as well as $\bL_0 - h$ and $\bbL_0 - \bh$, must send $\ket{\chi_{h,\bh}}$ into the submodule $\QuotMod{1,1} \otimes \QuotMod{1,1}$.  In order for the submodule generated by $\ket{\chi_{h,\bh}}$ to have trivial intersection with $\QuotMod{1,1} \otimes \QuotMod{1,1}$, it is therefore necessary and sufficient that $\ket{\chi_{h,\bh}}$ be a singular vector, for $\vir \oplus \vir$, in $\bcS'$.

So, consider first the singular vector of dimension $\brac{5,0}$ in $\VerMod{2} \otimes \VerMod{0}$.  Its lifts in $\bcS'$ have the form
\begin{equation}
\ket{\chi_{5,0}} = \brac{\bL_{-3} - \bL_{-2} \bL_{-1} + \frac{1}{6} \bL_{-1}^3} \ket{\bt} + \brac{a_1 \bL_{-5} + a_2 \bL_{-3} \bL_{-2}} \ket{\bvac},
\end{equation}
for some $a_1, a_2 \in \CC$, because $\VerMod{0} \otimes \VerMod{2}$ has no states of this conformal dimension.  These states are clearly annihilated by $\bbL_n$ with $n>0$ because $\bcS'$ has no states of antiholomorphic dimension $-1$.  Similarly, $\bbL_0 \ket{\chi_{5,0}} = 0$, because the dimension $\brac{5,0}$ singular descendant of $\ket{\bT}$ has already been set to zero in $\QuotMod{1,1} \otimes \QuotMod{1,1}$ (it is a descendant of $\bL_{-1} \ket{\bvac} = 0$).  The constraints coming from demanding that $\ket{\chi_{5,0}}$ be singular are therefore exactly the same as in the chiral case.  It is easy to check that this requires
\begin{equation}
a_1 = -\frac{1}{3}, \qquad a_2 = \frac{1}{2}, \qquad \bdb = \frac{5}{6}.
\end{equation}
In other words, we can consistently quotient $\bcS'$ by the submodule generated by $\ket{\chi_{5,0}}$ if and only if $\bdb = \tfrac{5}{6}$.  No constraint is imposed upon $\bbb$ in this analysis --- such a constraint would come from quotienting by the submodule generated by the appropriate lift $\ket{\bchi_{0,5}}$ of the singular vector in $\VerMod{0} \otimes \VerMod{2}$ of conformal dimension $\brac{0,5}$ (and would obviously require $\bbb = \tfrac{5}{6}$).

It should now be clear that the story regarding the singularity of $\ket{\chi_{7,0}}$ will also parallel the chiral case, leading thus to the well-known conclusion:  $\bdb = -\tfrac{5}{8}$.  Similarly, setting $\ket{\bchi_{0,7}}$ to zero requires $\bbb = -\tfrac{5}{8}$.  We therefore turn our attention to the singularity of $\ket{\chi_{2,1}}$.  Its form,
\begin{equation}
\ket{\chi_{2,1}} = \bbL_{-1} \ket{\bt},
\end{equation}
is particularly simple because neither $\QuotMod{1,1} \otimes \QuotMod{1,1}$ nor $\VerMod{0} \otimes \VerMod{2}$ have any dimension $\brac{2,1}$ states.  However, this state is never singular:
\begin{equation}
\bbL_1 \ket{\chi_{2,1}} = 2 \bbL_0 \ket{\bt} = 2 \ket{\bT}.
\end{equation}
We therefore cannot use it to derive interesting constraints on $\bdb$ or $\bbb$.  Note that it follows that the field $\func{\bt}{z,\bz}$ corresponding to $\ket{\bt}$ cannot be made holomorphic:  $\func{\bpa \bt}{z,\bz} \neq 0$.  Similarly, $\func{\bbt}{z,\bz}$ is not antiholomorphic.\footnote{One appealing consequence of this is that the module $\bcS'$ cannot be identified as a tensor product of two chiral staggered modules.}

Our last hope is therefore $\ket{\chi_{2,2}}$.  Here, things are more interesting because each of $\VerMod{2} \otimes \VerMod{0}$, $\QuotMod{1,1} \otimes \QuotMod{1,1}$ and $\VerMod{0} \otimes \VerMod{2}$ have states of dimension $\brac{2,2}$, though those of $\QuotMod{1,1} \otimes \QuotMod{1,1}$ are precisely the scalar multiples of the singular vector $\ket{\bT \bbT}$ of \eqnref{eqn:SVs}.  Taking the most general form for $\ket{\chi_{2,2}}$ modulo $\ket{\bT \bbT}$ and requiring annihilation under $\bL_1$, $\bbL_1$, $\bL_0 - 2$ and $\bbL_0 - 2$ fixes our candidate singular vector up to scalar multiples as
\begin{equation} \label{eqn:Chi22}
\ket{\chi_{2,2}} = \brac{\bbL_{-2} - \frac{3}{2} \bbL_{-1}^2} \ket{\bt} - \brac{\bL_{-2} - \frac{3}{2} \bL_{-1}^2} \ket{\bbt}.
\end{equation}
This has a suggestive interpretation:  The singular vector $\ket{\bT \bbT}$ has \emph{two} logarithmic partners in $\bcS'$, one lifted from $\VerMod{2} \otimes \VerMod{0}$ and the other from $\VerMod{0} \otimes \VerMod{2}$.  \eqnref{eqn:Chi22} gives $\ket{\chi_{2,2}}$ as the \emph{difference} of these lifts.  Setting $\ket{\chi_{2,2}}$ to zero would therefore amount to identifying these two logarithmic partner states.

It remains to impose annihilation under $\bL_2$ and $\bbL_2$.  An easy computation gives
\begin{equation} \label{eqn:Ann2}
\bL_2 \ket{\chi_{2,2}} = \brac{\bdb + 5} \ket{\bbT}, \qquad \bbL_2 \ket{\chi_{2,2}} = \brac{\bbb + 5} \ket{\bT},
\end{equation}
hence $\ket{\chi_{2,2}}$ is singular if and only if $\bdb = \bbb = -5$.  We therefore have a winner:  For $\bdb = \bbb = -5$, there is a non-chiral staggered module $\bcS = \bcS' / \bcT'$, where $\bcT'$ is the submodule generated by the singular vector $\ket{\chi_{2,2}}$ given in \eqref{eqn:Chi22}.  The existence of this module would appear to explain the observation of \cite{VasPuz11} completely.  We emphasise that we are not claiming that $\bcS$ \emph{is} the module investigated there, merely that we have found a structural interpretation for the value of $\bdb$ they measured.

The $\bdb = \bbb = -5$ module $\bcS$ that we have discovered still has infinitely many composition factors, so it is natural to look for further submodules that one can quotient by without affecting the essential non-chiral staggered structure.  We have shown above that one cannot set $\ket{\chi_{5,0}}$, $\ket{\chi_{7,0}}$, $\ket{\bchi_{0,5}}$ or $\ket{\bchi_{0,7}}$ to zero in $\bcS$ without conflicting with $\bdb = \bbb = -5$, but we expect that the constraints on the deeper singular vectors will not be as limiting.  Indeed, chiral considerations \cite{RidSta09} show that there is no obstacle to quotienting by the submodules generated by appropriately chosen $\ket{\chi_{12,0}}$, $\ket{\chi_{15,0}}$, $\ket{\bchi_{0,12}}$ and $\ket{\bchi_{0,15}}$.  Moreover, it is easy to check that the same is true for $\ket{\chi_{5,1}}$, $\ket{\chi_{7,1}}$, $\ket{\bchi_{1,5}}$ and $\ket{\bchi_{1,7}}$.  However, this is not the case for $\ket{\chi_{5,2}}$, $\ket{\chi_{7,2}}$, $\ket{\bchi_{2,5}}$ or $\ket{\bchi_{2,7}}$.  The structure diagram of the ``smallest'' $\bdb = \bbb = -5$ non-chiral staggered module $\mbcS$ is therefore as follows:
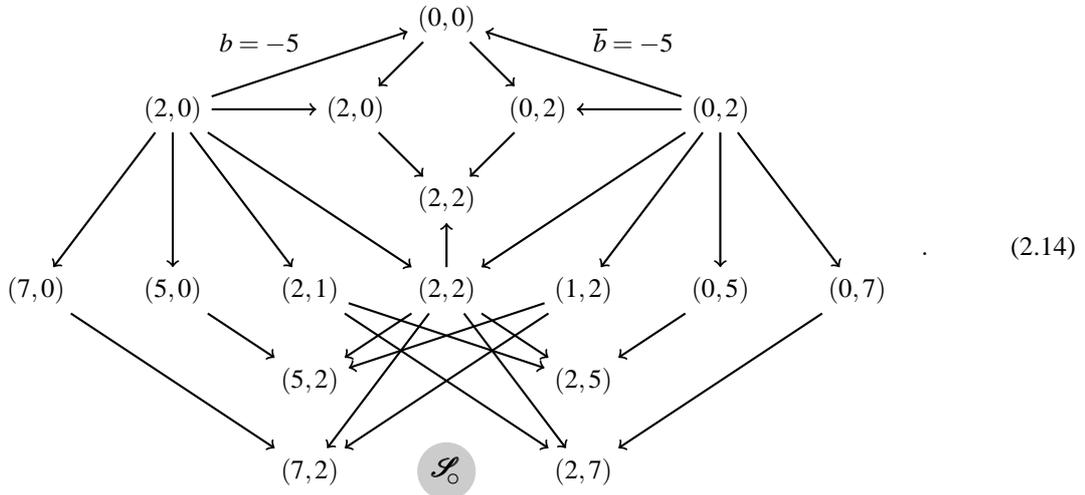
\begin{equation} \label{pic:S}
\parbox[c]{0.8\textwidth}{
\begin{center}
\begin{tikzpicture}[auto,thick,scale=0.6,
	nom/.style={circle,draw=black!20,fill=black!20,inner sep=2pt}
	]
\node (top) at (0,0) [] {$\brac{0,0}$};
\node (left) at (-2,-2) [] {$\brac{2,0}$};
\node (right) at (2,-2) [] {$\brac{0,2}$};
\node (bot) at (0,-4) [] {$\brac{2,2}$};
\node (l20) at (-6,-2) [] {$\brac{2,0}$};
\node (l21) at (-3,-6) [] {$\brac{2,1}$};
\node (l50) at (-6,-6) [] {$\brac{5,0}$};
\node (l70) at (-9,-6) [] {$\brac{7,0}$};
\node (m22) at (0,-6) [] {$\brac{2,2}$};
\node (m25) at (3,-8) [] {$\brac{2,5}$};
\node (m52) at (-3,-8) [] {$\brac{5,2}$};
\node (m27) at (3,-10) [] {$\brac{2,7}$};
\node (m72) at (-3,-10) [] {$\brac{7,2}$};
\node (r02) at (6,-2) [] {$\brac{0,2}$};
\node (r12) at (3,-6) [] {$\brac{1,2}$};
\node (r05) at (6,-6) [] {$\brac{0,5}$};
\node (r07) at (9,-6) [] {$\brac{0,7}$};
\node at (0,-10) [nom] {$\mbcS$};
\draw [->] (top) to (left);
\draw [->] (top) to (right);
\draw [->] (left) to (bot);
\draw [->] (right) to (bot);
\draw [->] (l20) to (left);
\draw [->] (l20) to (l21);
\draw [->] (l20) to (m22);
\draw [->] (l21) to (m25);
\draw [->] (l21) to (m27);
\draw [->] (l20) to (l50);
\draw [->] (l20) to (l70);
\draw [->] (l50) to (m52);
\draw [->] (l70) to (m72);
\draw [->] (m22) to (bot);
\draw [->] (m22) to (m25);
\draw [->] (m22) to (m27);
\draw [->] (m22) to (m52);
\draw [->] (m22) to (m72);
\draw [->] (r02) to (right);
\draw [->] (r02) to (r12);
\draw [->] (r02) to (m22);
\draw [->] (r12) to (m52);
\draw [->] (r12) to (m72);
\draw [->] (r02) to (r05);
\draw [->] (r02) to (r07);
\draw [->] (r05) to (m25);
\draw [->] (r07) to (m27);
\draw [->] (l20) to node {$\bdb = -5$} (top);
\draw [->] (r02) to node [swap] {$\bbb = -5$} (top);
\end{tikzpicture}
\end{center}
}
.
\end{equation}
We have again simplified the diagram by omitting arrows describing the action of the positive-mode subalgebra, except for those labelled by $\bdb$ and $\bbb$.  Nevertheless, this analysis makes it clear that the structure of bulk indecomposables is significantly more intricate than that of their boundary analogues.

It is tempting to claim that $\mbcS$ is the ``Ockham's razor'' candidate for the non-chiral staggered module that explains the measurement of \cite{VasPuz11}.  However, it is easy to check that this module is not physically satisfactory because it is not isomorphic to its contragredient dual (the other quotients of $\bcS$ suffer from this failing as well).  To restore non-degeneracy of the two-point functions, many of the ``deeper'' composition factors require additional logarithmic partners.  We therefore expect that $\mbcS$, or a similar quotient, will instead appear as a \emph{submodule} of the non-chiral module that appears in the bulk percolation (dilute polymers) model of \cite{VasPuz11}.  The physical module would then be some even larger indecomposable, presumably with infinitely many composition factors, in which the deeper structure is more complicated than our na\"{\i}ve rank $2$ considerations have allowed.  We will not try to address this structure here --- ``experimental'' input (that is, concrete examples) would seem to be a good idea at this point --- but hope to return to it in the future.

We conclude the section with a brief remark concerning the recent observation \cite{RunLog12} that the logarithmic coupling $\bdb = -5$ may also be recovered in a chiral setting when the $c=0$ vacuum $\ket{0}$ possesses a Jordan partner state $\ket{O}$ with $L_0 \ket{O} = \ket{0}$.\footnote{Actually, the chiral setup in \cite{RunLog12} involved a Jordan cell of rank $3$, but this is irrelevant to the argument.}  For
\begin{equation}
L_2 \brac{L_{-2} - \frac{3}{2} L_{-1}^2} \ket{O} = -5 L_0 \ket{O} = -5 \ket{0}.
\end{equation}
Comparing with \eqnDref{eqn:Chi22}{eqn:Ann2} reveals that this is not really a coincidence.  It is in fact a general phenomenon, if not a particularly deep one.  Nevertheless, we emphasise that the measurement of \cite{VasPuz11} did not pertain to a chiral module, and more importantly, no Jordan partners to the vacuum were reported there.

\section{Other Non-Chiral Staggered Modules} \label{sec:Other}

It should be clear that the above analysis is not restricted to extending the non-chiral vacuum module $\QuotMod{1,1} \otimes \QuotMod{1,1}$, nor to $c=0$.  Here, we outline the complementary $c=0$ case in which $\bcS'$ has submodule $\QuotMod{1,2} \otimes \QuotMod{1,2}$, with $\QuotMod{1,2} = \VerMod{0} / \VerMod{2}$, and its quotient by this submodule is $\brac{\VerMod{1} \otimes \VerMod{0}} \oplus \brac{\VerMod{0} \otimes \VerMod{1}}$.  We let $\ket{\bOmega}$ denote the \hws{} of $\QuotMod{1,2} \otimes \QuotMod{1,2}$ and denote the logarithmic partners of $\bL_{-1} \ket{\bOmega}$ and $\bbL_{-1} \ket{\bOmega}$ by $\ket{\bomega}$ and $\ket{\bbomega}$, respectively:
\begin{equation}
\brac{\bL_0 - 1} \ket{\bomega} = \bbL_0 \ket{\bomega} = \bL_{-1} \ket{\bOmega}, \qquad \brac{\bbL_0 - 1} \ket{\bbomega} = \bL_0 \ket{\bbomega} = \bbL_{-1} \ket{\bOmega}.
\end{equation}
The logarithmic couplings are defined by $\bL_1 \ket{\bomega} = \beta \ket{\bOmega}$ and $\bbL_1 \ket{\bbomega} = \bbeta \ket{\bOmega}$.

The analysis is even easier than that of the previous section.  The singular vectors of $\VerMod{1} \otimes \VerMod{0}$ which generate its maximal proper submodule have conformal dimensions $\brac{5,0}$, $\brac{7,0}$, $\brac{1,1}$ and $\brac{1,2}$.  Those of dimension $\brac{5,0}$ or $\brac{7,0}$ merely reproduce the chiral analysis upon lifting to $\bcS'$:  They become singular, hence may be set to zero, if and only if $\beta = -\tfrac{1}{2}$ or $\beta = \tfrac{1}{3}$, respectively.  It is likewise easy to check that the dimension $\brac{1,2}$ singular vector never lifts to a singular vector in $\bcS'$ because applying $\bbL_2$ to any such lift gives a non-zero multiple of $\bL_{-1} \ket{\bOmega}$.  The story for $\bbeta$ and $\VerMod{0} \otimes \VerMod{1}$ is identical (with holomorphic and antiholomorphic dimensions exchanged).

The truly non-chiral phenomenon is again found with the two Jordan partners to the dimension $\brac{1,1}$ singular vector $\bL_{-1} \bbL_{-1} \ket{\bOmega}$ of $\QuotMod{1,2} \otimes \QuotMod{1,2}$.  Taking their difference,
\begin{equation}
\ket{\chi_{1,1}} = \bbL_{-1} \ket{\bomega} - \bL_{-1} \ket{\bbomega},
\end{equation}
gives the only candidate for (another) dimension $\brac{1,1}$ singular vector in $\bcS'$, and it is trivial to check that $\ket{\chi_{1,1}}$ is singular if and only if $\beta = \bbeta = 2$.  We conclude from this that when $\beta = \bbeta = 2$, $\ket{\chi_{1,1}}$ may be set to zero to obtain a quotient module $\bcS$.  In any theory incorporating this module, the fields $\func{\bomega}{z,\bz}$ and $\func{\bbomega}{z,\bz}$ are neither holomorphic nor antiholomorphic, but satisfy
\begin{equation}
\func{\bpa \bomega}{z,\bz} = \func{\pa \bbomega}{z,\bz}.
\end{equation}

The obvious generalisation of this setup is as follows.  Let $\QuotMod{}$ and $\ahol{\QuotMod{}}$ be highest weight $\vir$-modules of length $2$, meaning that they have precisely two composition factors, and arbitrary central charge $c$.  Let these composition factors be $\IrrMod{h}$ and $\IrrMod{h+n}$ for $\QuotMod{}$ and $\IrrMod{\bh}$ and $\IrrMod{\bh+\bn}$ for $\ahol{\QuotMod{}}$.\footnote{We will consider modules of arbitrary (but fixed) central charge $c$ for the remainder of this section.}  The tensor product $\QuotMod{} \otimes \ahol{\QuotMod{}}$ will therefore have four singular vectors which we shall denote by $\ket{\bx}$, $\bU \ket{\bx}$, $\bbU \ket{\bx}$ and $\bU \bbU \ket{\bx}$ ($\bU$ and $\bbU$ are therefore linear combinations of monomials in the $\bL_n$ and $\bbL_n$, respectively).  There is then a two-parameter family of non-chiral staggered modules with submodule $\QuotMod{} \otimes \ahol{\QuotMod{}}$ and quotient $\brac{\VerMod{h+n} \otimes \VerMod{\bh}} \oplus \brac{\VerMod{h} \otimes \VerMod{\bh + \bn}}$.  We denote our choice of Jordan partners of $\bU \ket{\bx}$ and $\bbU \ket{\bx}$ by $\ket{\by}$ and $\ket{\bby}$, respectively:
\begin{equation}
\brac{\bL_0 - h - n} \ket{\by} = \brac{\bbL_0 - \bh} \ket{\by} = \bU \ket{\bx}, \qquad \brac{\bbL_0 - \bh - \bn} \ket{\bby} = \brac{\bL_0 - h} \ket{\bby} = \bbU \ket{\bx}.
\end{equation}
The logarithmic couplings are therefore $\beta = \bracket{\bx}{\bU^{\dag}}{\by}$ and $\bbeta = \bracket{\bx}{\bbU^{\dag}}{\bby}$.  We may represent the non-chiral staggered structure thus:
\begin{equation}
\parbox[c]{0.53\textwidth}{
\begin{center}
\begin{tikzpicture}[auto,thick,scale=0.6]
\node (top) at (0,0) [] {$\ket{\bx}$};
\node (left) at (-2,-2) [] {$\bU \ket{\bx}$};
\node (right) at (2,-2) [] {$\bbU \ket{\bx}$};
\node (bot) at (0,-4) [] {$\bU \bbU \ket{\bx}$};
\node (jleft) at (-6,-2) [] {$\ket{\by}$};
\node (jbot1) at (-4,-4) [] {$\bbU \ket{\by}$};
\node (jright) at (6,-2) [] {$\ket{\bby}$};
\node (jbot2) at (4,-4) [] {$\bU \ket{\bby}$};
\draw [->] (top) to (left);
\draw [->] (top) to (right);
\draw [->] (left) to (bot);
\draw [->] (right) to (bot);
\draw [->] (jleft) to (left);
\draw [->] (jleft) to (jbot1);
\draw [->] (jbot1) to (bot);
\draw [->] (jright) to (right);
\draw [->] (jright) to (jbot2);
\draw [->] (jbot2) to (bot);
\draw [->] (jleft) to node {$\beta$} (top);
\draw [->] (jright) to node [swap] {$\bbeta$} (top);
\end{tikzpicture}
\end{center}
}
.
\end{equation}

Note that the singular vector $\bU \bbU \ket{x}$ has two Jordan partners, $\bbU \ket{\by}$ and $\bU \ket{\bby}$.  We will therefore look for a singular vector of the form
\begin{equation}
\ket{\chi} = \bbU \ket{\by} - \bU \ket{\bby} + \bV \bbV \ket{x},
\end{equation}
where $\bV$ is a linear combination of grade $n$ monomials in the $\bL_m$ and $\bbV$ is a linear combination of grade $\bn$ monomials in the $\bbL_m$.  It is easily checked that $\bL_0 - h - n$ and $\bbL_0 - \bh - \bn$ both annihilate $\ket{\chi}$.  Moreover, a generalisation of the analysis of \cite[Sec.~7]{RidSta09} allows us to conclude that there will exist a choice of $\bV$ and $\bbV$ making $\ket{\chi}$ singular if and only if $\bU^{\dag} \ket{\bchi} = \bbU^{\dag} \ket{\bchi} = 0$.  Since $\bU^{\dag} \bV \ket{\bx} = \bracket{\bx}{\bU^{\dag} \bV}{\bx} \ket{\bx} = 0$ by the singularity of $\bU \ket{\bx}$ (and similarly, $\bbU^{\dag} \bbV \ket{\bx} = 0$), we obtain
\begin{equation}
\bU^{\dag} \ket{\chi} = \brac{\beta - \func{p}{\bL_0}} \bbU \ket{\bx}, \qquad \bbU^{\dag} \ket{\chi} = \brac{\bbeta - \func{\bp}{\bbL_0}} \bU \ket{\bx}.
\end{equation}
Here, we have used the fact that $\bU^{\dag} \bU$ annihilates $\ket{\bx}$ to write $\bU^{\dag} \bU = \func{p}{\bL_0} \brac{\bL_0 - h} + \bU_1 \bL_1 + \bU_2 \bL_2$, where $p$ is a polynomial in $\bL_0$ (and similarly for $\bbU^{\dag} \bbU$ and $\func{\bp}{\bbL_0}$).  From this, we conclude that $\ket{\chi}$ is singular, hence may be set to zero, precisely when
\begin{equation} \label{eq:GenLogCoup}
\beta = \func{p}{h}, \qquad \bbeta = \func{\bp}{\bh}.
\end{equation}

In the case treated above with $h = \bh = 0$ and $n = \bn = 1$, we have $\bU = \bL_{-1}$, hence $\bU^{\dag} \bU = 2 \bL_0 + \bL_{-1} \bL_1$ and $\func{p}{\bL_0} = 2$.  Similarly, $\func{\bp}{\bbL_0}$ is also the constant polynomial $2$, and we recover our conclusion that the non-chiral staggered module exists when $\beta = \bbeta = 2$.  The case treated in the previous section corresponds to $h = \bh = 0$ and $n = \bn = 2$.  Now, $\bU = \bL_{-2} - \tfrac{3}{2} \bL_{-1}^2$ and one obtains $\func{p}{\bL_0} = 18 \bL_0 - 5$.  The existence of the non-chiral staggered module therefore requires $\bdb = \beta = \func{p}{0} = -5$ and we recover $\bbb = \bbeta = -5$ in exactly the same manner.

\section{Concluding Remarks}

In this note, we have used the representation theory of $\vir \oplus \vir$ to construct \emph{non-chiral staggered modules} which are consistent with the recent measurement \cite{VasPuz11} of a non-chiral logarithmic coupling $\beta = \bdb = -5$ for a $c=0$ indecomposable containing the vacuum $\ket{\bvac} = \ket{0} \otimes \ket{0}$.  More precisely, we have shown that there exist non-chiral staggered modules whose structure requires that $\bdb = \bbb = -5$ and, moreover, that this is the \emph{only} such value for the logarithmic couplings, beyond the chiral values $\tfrac{5}{6}$ and $-\tfrac{5}{8}$, whose origin has a structural explanation.  We are therefore confident in predicting that the indecomposable module studied in \cite{VasPuz11} will contain one of our non-chiral staggered modules as a submodule.  It is clear that this inclusion must be proper, as the predicted modules are not self-contragredient on their own.

The fact that the value $\bdb = -5$ is not forbidden by the representation theory suggests reconsidering the conclusion of \cite{VasPuz11} that the four point function of the ``energy operator'', corresponding to the \hws{} of $\IrrMod{5/8}$ for percolation and $\IrrMod{1/3}$ for dilute polymers, must be set to zero.  The argument is that the relevant $c=0$ conformal blocks lead to the chiral values $\bdb = -\tfrac{5}{8}$ or $\tfrac{5}{6}$ \cite{GurCon02}.  However, this ignores the fact that the blocks must be combined in a bulk theory to get local correlation functions.  From the viewpoint of fusion, this means that the na\"{\i}ve \emph{non-chiral} fusion rules must be modified appropriately.  In particular, one should quotient \cite{GabLoc99} the following guesses by the image of the nilpotent part of the spin operator $\bL_0 - \bbL_0$:
\begin{equation}
\begin{split}
\brac{\IrrMod{5/8} \otimes \IrrMod{5/8}} \fuse \brac{\IrrMod{5/8} \otimes \IrrMod{5/8}} &= \StagMod{\bdb = 5/6} \otimes \StagMod{\bdb = 5/6}, \\
\brac{\IrrMod{1/3} \otimes \IrrMod{1/3}} \fuse \brac{\IrrMod{1/3} \otimes \IrrMod{1/3}} &= \StagMod{\bdb = -5/8} \otimes \StagMod{\bdb = -5/8}.
\end{split}
\end{equation}
Here, $\StagMod{\bdb = 5/6}$ and $\StagMod{\bdb = -5/8}$ denote the chiral $c=0$ extensions of $\QuotMod{3,1}$ and $\QuotMod{1,5}$, respectively, by $\QuotMod{1,1}$, with the given (chiral) logarithmic couplings.  Performing the quotients, one finds that the resulting non-chiral modules contain holomorphic and antiholomorphic energy-momentum tensors which lack logarithmic partners.  We therefore conclude that the argument of \cite{GurCon02} is inapplicable to bulk considerations.  What this analysis does suggest, however, is that the energy operators do not appear in the bulk $\bdb = -5$ theory as irreducible tensor irreducible; indeed, this is not the structure seen on the lattice \cite{VasPuz11}.  It would be interesting to know if $\IrrMod{-1/24}$ and $\IrrMod{35/24}$ appear in the bulk with the simple irreducible tensor irreducible structure.

We have also generalised our construction to arbitrary central charges and explained how to compute the logarithmic couplings in the general case (\eqnref{eq:GenLogCoup}).  One can therefore calculate these bulk couplings explicitly as functions of the central charge, at least in principle.  If we take $h = h_{r,s}$ ($r,s \in \ZZ_+$) and $n = rs$ in the general setting of \secref{sec:Other}, where
\begin{equation}
h_{r,s} = \frac{r^2-1}{4} t - \frac{rs-1}{2} + \frac{s^2-1}{4} t^{-1}, \qquad c = 13 - 6 \brac{t + t^{-1}}
\end{equation}
as usual, then we have been able to determine the non-chiral logarithmic coupling $\beta \equiv \beta_{r,s}$ explicitly when $rs \leqslant 8$.  In particular, the $\beta_{1,s}$ thus computed are consistent with the following conjecture (see also \cite[Ex.~16]{RidSta09}):
\begin{equation} \label{conj:Non-Chiral}
\beta_{1,s} = \frac{2 \brac{-1}^{s-1} s! \brac{s-1}!}{t^{2 \brac{s-1}}} \prod_{j=1}^{s-1} \brac{t^2 - j^2}.
\end{equation}
Here, we have assumed, in contrast to previous sections, the \emph{first} normalisation convention of \eqref{eqn:NormSV}.  We remark that $\beta_{s,1}$ may be obtained from this formula by inverting $t$.

We compare this with the chiral logarithmic couplings, extending the results (and conjectures) of \cite{RidLog07,KytFro08}.  For any value of the central charge, there are two distinguished chiral couplings, $\beta_{1,s}^{\rightarrow}$ and $\beta_{1,s}^{\downarrow}$, which arise, for example, when explicitly computing the (respective) fusion rules\footnote{We remark that there is no requirement for $p$ and $q$ to be coprime integers for these computations.}
\begin{equation}
\QuotMod{1,s+1} \fuse \VerMod{p-1,q}, \qquad \QuotMod{2,1} \fuse \VerMod{p,q-s} \qquad \text{($t = q/p$).}
\end{equation}
We have performed this computation for $s \leqslant 4$ and the results are consistent with the conjectures (see also \cite{RidLog07})
\begin{equation} \label{conj:Chiral}
\begin{split}
\beta_{1,s}^{\rightarrow} &= \frac{\brac{-1}^{s-1} s! \brac{s-1}!}{t^{2 \brac{s-1}}} \prod_{j=1}^{s-1} \brac{t^2 - j^2} \cdot \frac{t-s}{-s}, \\
\beta_{1,s}^{\downarrow} &= \frac{\brac{-1}^{s-1} s! \brac{s-1}!}{t^{2 \brac{s-1}}} \prod_{j=1}^{s-1} \brac{t^2 - j^2} \cdot \frac{t-s}{t}.
\end{split}
\end{equation}
One can perform similar computations with $r \geqslant 1$, but the computational intensity of the fusion algorithm severely limits our ability to make more general conjectures.  We expect that the heuristic methods reported in \cite{VasInd11} for computing logarithmic couplings will be far better suited for this purpose.

Recall that \cite{VasPuz11} remarked that the inverse of the non-chiral $c=\ahol{c}=0$ coupling $\beta = -5$ is the average of the inverses of the chiral couplings $-\tfrac{5}{8}$ and $\tfrac{5}{6}$.  This coincidence is presumably backed up by many other examples.  The general computations reported above completely confirm that this is a general phenomenon.  In particular, the conjectures \eqref{conj:Non-Chiral} and \eqref{conj:Chiral} satisfy
\begin{equation}
\beta_{1,s}^{-1} = \frac{1}{2} \sqbrac{\brac{\beta_{1,s}^{\rightarrow}}^{-1} + \brac{\beta_{1,s}^{\downarrow}}^{-1}},
\end{equation}
for all central charges.  We have also checked this observation for several other (small) values of $r$ and $s$ and for all $t$.  This surprising relation between the chiral and non-chiral logarithmic couplings suggests a deeper relation between the singular vectors which control their values.  Understanding why this relation should hold (and indeed, why there should be any relation at all!) seems to us to be an important outstanding puzzle in the representation theory of the Virasoro algebra.

We would like to conclude by emphasising two aspects of the results reported here.  First, the above analysis makes it clear that the representation theory does not uniquely pin down non-chiral logarithmic couplings.  Indeed, it is only in a few cases that setting singular vectors to zero constrains the otherwise arbitrary couplings.  This should be familiar from the analogous chiral analyses, but it is relevant given the claim in \cite{VasPuz11} that their (heuristic) theoretical methods pick out $\bdb = -5$ uniquely.  This could be argued to be a consequence of design:  In the \ope{} approach, the divergences of the expansion of a non-chiral field with itself as $c$ and $\ahol{c}$ tend to $0$ were cancelled by introducing a \emph{single} field $\func{X}{z,\ahol{z}}$ whose properties lead, inexorably, to $\bdb = -5$.  We propose that the divergence could also have been cancelled using two separate fields, one for $c \to 0$ and the other for $\ahol{c} \to 0$, and that doing so would lead to the familiar chiral values for $\bdb$ (and $\bbb$).  Similarly, the Coulomb gas approach likewise makes use of $\func{X}{z,\ahol{z}}$ and its properties, so it is perhaps not surprising that this technique also recovers $\bdb = -5$.  That said, using a single field is certainly very natural.  We find our identification of two logarithmic partners to the same state to be likewise very natural.

The second aspect that we wish to emphasise is that the general results reported above are contingent upon the Ansatz that the submodule $\QuotMod{} \otimes \ahol{\QuotMod{}}$ of the non-chiral staggered module is the tensor product of two indecomposable $\vir$-modules of length $2$.  This means that one must introduce two independent logarithmic partner states.  Obviously, other Ans\"{a}tze are possible.  In particular, one expects that at $c=-2$, the vacuum generates an irreducible $\vir$-module $\IrrMod{0}$, so that it is natural to introduce only one logarithmic partner state.  The analysis is therefore significantly easier than that reported here, but is still interesting because in this case, one has clear predictions from the theory of symplectic fermions \cite{GabLoc99} and abelian sandpiles \cite{JenHei06} for the physically relevant non-chiral structures.  We expect that understanding this in detail will shed light on the problem of constructing self-dual non-chiral staggered modules at $c=0$ and more generally.

\section*{Acknowledgements}

I would like to thank Romain Vasseur for several discussions pertaining to $c=0$ bulk indecomposables, thereby initiating this study.  I also thank Philippe Ruelle and Ingo Runkel for valuable correspondence on this matter and Hubert Saleur for commenting on a draft manuscript.  This work was begun at the Institut Henri Poincar\'{e} during the recent trimestre ``Advanced Conformal Field Theory and Applications''.  I am grateful to the organisers and institute staff for their gracious hospitality.  This research was supported by the Australian Research Council Discovery Project DP0193910.

\raggedright

\end{document}